\documentclass[12pt,twoside]{article}
\usepackage{fleqn,espcrc1}

\usepackage{graphicx}
\newcommand{\AmS}{{\protect\the\textfont2
  A\kern-.1667em\lower.5ex\hbox{M}\kern-.125emS}}

\title{ A test of the Kugo-Ojima confinement criterion by lattice Landau gauge QCD simulations}

\author{Hideo Nakajima\thanks{e-mail nakajima@is.utsunomiya-u.ac.jp}\\ 
Department of Information science, Utsunomiya University  \\
Sadataka Furui \thanks{e-mail furui@dream.ics.teikyo-u.ac.jp}\\
School of Science and Engineering, Teikyo University}

\begin{document}
% typeset front matter
\def\today{Feb. 9. 2000}
\maketitle

\begin{abstract}
The first test of the Kugo-Ojima colour confinement criterion by
the lattice Landau gauge QCD simulation is performed.
The parameter $u$ which is expected to be $-1\delta^a_b$ in the
continuum theory was found to be $-0.7\delta^a_b$ in the strong
coupling region.  The data is analysed in connection with the theory of 
Zwanziger. 
In the weak coupling region, the expectation value of the horizon function
 is negative or consistent to 0.
\end{abstract}

\section{Introduction}
There are various manifestation of colour confinement in QCD. One is the linear
potential between quarks, which appears in the quenched lattice QCD simulation. 
A mechanism for the appearance of the linear potential was proposed as well by 
Gribov
\cite{Gv} about 20 years ago. He showed that the Landau gauge fixing or the 
Coulomb gauge fixing does not fix the gauge field $A_\mu$ uniquely and that
the restriction of $A_\mu$ to a physical space which is called the Gribov 
region enhances the singularity in the ghost propagator and induces the
linear potential. 

In the field theory, the confinement implies absence of free single coloured 
particle state in the asymptotic Hilbert space.  The physical Hilbert space 
should satisfy symmetry specified by the Lagrangian, and the QCD 
Lagrangian is invariant under the BRS(Becchi-Rouet-Stora) transformation.
Kugo and Ojima\cite{KO} proposed in 1978, a criterion for the 
colour confinement in the Landau gauge, based on the BRS symmetry, which 
consists  of a two-point function produced by 
the ghost, the antighost and the gauge field becomes $-\delta_a^b$, where 
$a$ and $b$ specify the colour in the adjoint representation. Analytical 
calculation of this value is extremely difficult and so far no verification 
was performed.

In the lattice QCD, the Gribov region still
does not define the gauge field uniquely but there is a unique minimum $\|A\|$
in the fundamental modular region\cite{Zw}.  He argued that the restriction
to the fundamental modular region implies the regularity of the horizon tensor,
whose transverse projection is identical to the $-$ of the two-point function of Kugo-Ojima. 

Both the Gribov-Zwanziger's theory and the Kugo-Ojima's theory suggest that in 
the lattice Landau gauge, the gluon propagator is infrared finite, which is
confirmed by the lattice QCD simulation\cite{MO,NF,LSWP}. 
In this paper we simulate the two-point function specified by the Kugo-Ojima  
in the lattice Landau gauge.

\section{The Kugo-Ojima confinement criterion and the Gribov-Zwanziger's theory}
\subsection{Kugo-Ojima's theory}

A sufficient condition of the colour confinement given 
by Kugo and Ojima\cite{KO} is that $u^a_b$ defined by the two-point
function of  the FP (Faddeev-Popov) ghost fields, $c(x),\bar c(y)$,
and $A_\nu(y)$,
\begin{equation}
\int e^{ip(x-y)}\langle 0|T D_\mu c^a(x)g(A_\nu\times \bar c)_b(y)|0\rangle dx=(g_{\mu\nu}-{p_\mu p_\nu\over p^2})u^a_b(p^2)
\label{eq}
\end{equation}
satisfies $u^a_b(0)=-\delta^a_b$.

Essential points in their argument is based on the BRS invariance of the QCD 
Lagrangian accompanied by the gauge fixing and FP terms.
  1) The Ward-Takahashi identities implies that the gauge fields
$A^a_\mu(x)$, the auxiliary field $B^a$, the covariant derivative of the
ghost field $D_\mu c^a$ and the antighost field $\bar c^a(x)$ necessarily have
massless asymptotic fields which forms the BRS-quartet.
The Hilbert space is decomposed into the FP ghost number eigenstates and
in this bases, the BRS-quartet space have the zero-norm.  
2) The BRS charge $Q_B$ is conserved, and
the Noether current corresponding to the conservation of the
colour symmetry is
\begin{equation}
gJ^a_\mu={\partial ^\nu}{F^a_{\mu\nu}}+\{Q_B, D_\mu \bar c\},
\end{equation}
where its ambiguity by divergence of antisymmetric tensor should be understood,
and this ambiguity is utilised so that massless contribution may be eliminated
for the charge, $Q^a$, to be well defined.
The massless component in the current $\{Q_B, D_\mu \bar c\}$ is absent if
 ${\bf 1+u=0}$.

\subsection{Gribov-Zwanziger's theory}

The Landau gauge of the QCD is specified by $\partial_\mu A_\mu=0$.  
The Gribov region $\Omega$ is specified by
ensemble of local minimum points of gauge orbits under the variation with 
respect to
$g=e^\epsilon$ as follows.
\begin{equation}
\Delta\|A^g\|^2=-2\langle \partial A|\epsilon\rangle+
\langle \epsilon|-\partial {\cal D}|\epsilon\rangle+\cdots
\label{NORM6}
\end{equation}
\begin{equation}
\Omega=\{A|-\partial {\cal D}\ge 0\ ,\ \partial A=0\}\ \ .
\label{NORM8}
\end{equation}

The physical space of the gauge field is characterized by the condition
that the FP determinant is positive. In the Coulomb gauge, 
the singulatity of the ghost propagator yields enhancement of the infrared 
singularity of the Coulomb potential\cite{Gv}. 

In the lattice simulation, the unique gauge field configuration can be
attained by the restriction to the fundamental modular region $\Lambda$\cite{Zw}, which is specified by the
absolute minimum along the gauge orbits.
\begin{equation}
\Lambda=\{A|\|A\|^2={\rm Min}_g\|A^g\|^2\}, \qquad
\Lambda\subset \Omega\ \ .
\label{NORM9}
\end{equation}

Let the gauge configuration be in the fundamental modular region
obtained by the optimising function $\displaystyle I(U^g)=\sum_{x,\mu}(1-{1\over n}Re tr U^g_{x,\mu})$, and let it be a global minimum even under the gauge transformation of larger period (the region is called the core region),
and the two point tensor be defined as
${G_{\mu\nu xy}}\delta^{ab}
= \langle{\rm tr}\left({\lambda^a}^{\dag}
D_\mu \displaystyle{1\over -\partial D}(-D_\nu)\lambda^b\right)_{xy}\rangle$.
 Then, in the Zwanziger's theory\cite{Zw,Cu}, the horizon function $H(U)$ 
 defined as
\begin{equation}
\displaystyle{\langle H(U)\rangle\over V}
=(N^2-1)\left[\lim_{p\to 0}
G_{\mu\mu}(p)-e\right]
\label{INFZW}
\end{equation}
is negative or 0 in finite volume, and 0 in the infinite volume limit.
Here $N$ is the number of colour, $V$ is the lattice volume, 
and $\displaystyle e=\langle \sum_l{1\over N}Re tr U_l \rangle/V$.

Note that in a $d$ dimensional lattice $e/d=1$ if all links $U_l=1$, 
and that the value of $1-e/d$ has the meaning of the distance from this 
vacuum.   

\section{The Lattice simulation}

We define the gauge field\cite{NF}
on links as an element of $SU(3)$ Lie algebra as,
and perform the gauge transformation as
$e^{A^g_{x,\mu}}=g_x^\dagger  e^{A_{x,\mu}} g_{x+\mu}$

The Landau gauge is realised by minimising $\|A^g\|^2$ via a gauge
transformation $g^\dagger U g$, where $g=e^\epsilon$. $\epsilon$ is obtained by solving the equation with a suitable parameter $\eta$
\begin{equation}
{\cal M}\epsilon=-\partial_\mu D_\mu(A) \epsilon=\eta\partial A 
\end{equation}

The obtained norm $\|A \|$ is close to that 
obtained after the smeared gauge fixing\cite{HdF} within 1\%.

 The inverse FP operator, ${\cal M}^{-1}[U]=(M_0-M_1[U])^{-1}$, 
is calculated
 perturbatively by using the Green function of the  Poisson equation\cite{NF}.

 In use of colour source $|\lambda^a x\rangle$ normalised as
$tr \langle \lambda^a x|\lambda^b x_0\rangle=\delta^{ab}\delta_{x,x_0}$,
the ghost propagator is given by
\begin{equation}
G^{ab}(x,y)=\langle tr \langle \lambda^a x|({\cal M}[U])^{-1}|
\lambda^b y\rangle \rangle
\end{equation}
where the outmost $\langle\rangle$
specifies average over samples $U$. 

The ghost propagator is infrared divergent and its singularity can be
parametrised as $\tilde Z_3 p^{-2-\alpha}=\tilde Z_3(4 \sin^2{ \pi\over L})^{-1-\alpha/2}$. 
The data of $\beta=5.5$, $8^3\times 16$ 
lattice indicates that the singularity is approximately $p^{-2.2}$, 
while the data of $16^4$ lattice indicates that it is $p^{-2.8}$.  The finite-size effect is not so large\cite{SS}.
These qualitative features are in agreement with the 
analysis of the Dyson-Schwinger equation\cite{SHA}.

Using the above inverse FP operator, we obtained that $u^a_b(0)$ is consistent to $-c\delta^a_b,\ c=0.7$ in
$SU(3)$ quenched simulation, $\beta =5.5$, on $8^4, 12^4$ and $16^4$. Fig.2
 shows the value of $|u^a_a|$.
 
\begin{figure}[htb]
\begin{minipage}[b]{0.47\linewidth} 
\begin{center}
\includegraphics[scale=0.6]{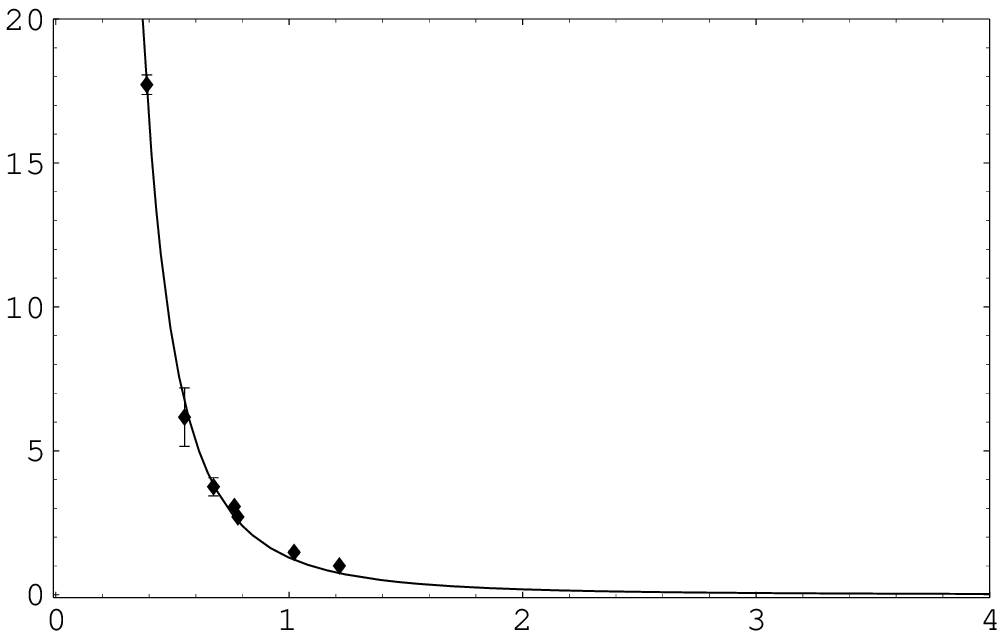}
\caption{The ghost propagator as function of the lattice momentum. 
The data are $\beta=5.5, 16^4$. The fitted curve is $1.287/ p^{2.779}$.}
\end{center}
\end{minipage}
\hfil
\begin{minipage}[b]{0.47\linewidth} 
\begin{center}
\includegraphics[scale=0.6]{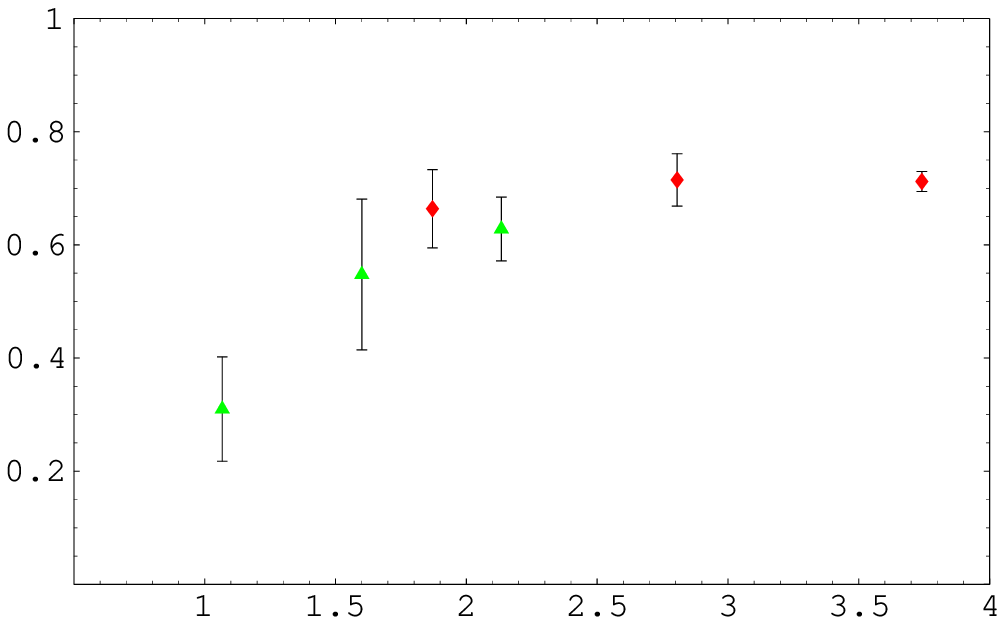}
\caption{The Kugo-Ojima parameter $|u^a_a|$ as the function of the spatial 
extent of the lattice $a L (fm)$.  The data are $\beta=6, 8^4, 12^4, 16^4$ and $\beta=5.5, 8^4, 12^4, 16^4$ from left to right, respectively.}
\end{center}\label{kugo1}
\end{minipage}
\end{figure}

In terms of the Kugo-Ojima parameter $c$, the function $G_{\mu\mu}(0)$ can be 
written as $1+3c$. 
In the $16^4$ lattice,  $e/d=0.78$ and $0.86$ for $\beta=5.5$ and $\beta=6$, respectively. They are numerically close to the inverse of the renormalization factor of
the ghost propagator $1/\tilde Z_3$. The Slavnov-Taylor relation 
$\displaystyle Z_1={Z_3/\tilde Z_3}=1$ implies $e/d \simeq 1/Z_3$. 

\begin{table}[th]
\begin{tabular*}{\textwidth}{@{}l@{\extracolsep{\fill}}cccccccc}
%\caption{$\beta$ dependence of the Kugo-Ojima parameter $c$ and the 
%renormalization factor $1/\tilde Z_3$.} 
\hline
%INSERT HERE THE DATA FILE.....START:
    & $\beta$    & c  &   $G_{\mu\mu}(0)$ & $e$ & $e/d$ &${\cal S}$& $1/\tilde Z_3$\\
\hline
 &5.5  & 0.712(18)  &  3.14(5)  & 3.13(1) & 0.783 &0.657 & 0.777\\
 &6.0  & 0.628(56)  &  2.88(17)  & 3.45(1)& 0.863 &0.694 & 0.860\\
\hline
\end{tabular*}
\end{table}

Our data of $16^4$ lattice suggest that when $\beta$ becomes 
larger, $G_{\mu\mu}(0)$
becomes smaller, while the Zwanziger's parameter $e$ has the opposite tendency.

It is to be remarked that in the Zwanziger's theory, Kugo-Ojima criterion $c=1$
 does not hold in view of $e/d \ne 1$. 
In the case of $\beta=5.5$, our data of $G_{\mu\mu}(0)$ agrees 
with $e$. However,  when the optimising function $\displaystyle I(A^g)=\| A \|^2=\sum_{x,\mu}tr {A^g}^\dagger A^g$ is used instead of $I(U^g)$, the function $e/d$ is to be
replaced by the link average
$\langle {\cal S}({\cal A})\rangle =\langle {{\cal A}/2\over \tanh {\cal A}/2}
\rangle$, where ${\cal A}=adj A$, and the value is reduced by about 20\%. 
The positive horizon function for $\beta=5.5$ implies our configurations are 
not in the core region. 
In the weak coupling region ($\beta=6.0$), the horizon function is negative or
consistent to 0.

\end{document}